\documentclass[11pt]{amsart}
\usepackage{epsfig}
\begin{document}
\topmargin=5mm
 \oddsidemargin=-2mm
 \evensidemargin=-2mm
\baselineskip=20pt
\parindent 20pt
{\flushleft{\Large\bf  A Riemann theta function formula with
its application to  double periodic wave solutions of nonlinear  equations }}\\[12pt]
{\large\bf  Engui Fan$^{a}$\footnote{\ \  E-mail
address: \  faneg@fudan.edu.cn} and Kwok Wing Chow$^{b}$ }\\[8pt]
{\small a. School of Mathematical Sciences and Key Laboratory of
Mathematics for Nonlinear Science,\\ Fudan University, Shanghai,
200433, P.R.  China\\
b. Department of Mechanical Engineering, University of Hong Kong,
Pokfulam, Hong Kong} \vspace{6mm}
\begin{center}
\begin{minipage}{5in}
\baselineskip=17pt { \small {\bf Abstract.}  Based on a Riemann
theta function and Hirota's bilinear form,  a lucid and
straightforward way is presented to  explicitly construct double
periodic wave solutions  for both nonlinear differential and
difference equations. Once such a equation is written in a bilinear
form, its periodic wave solutions can be directly obtained by using
an unified theta function formula. The relations between the
periodic wave solutions and soliton solutions are rigorously
established.  The efficiency of our proposed method can be
demonstrated on a class variety of nonlinear equations such as those
considered in this paper, shall water wave equation,
(2+1)-dimensional Bogoyavlenskii-Schiff
equation and differential-difference KdV equation. 
}
\end{minipage}\\[12pt]
\end{center}
{\bf\large 1. Introduction}\\

The bilinear derivative method developed by Hirota is a powerful and
direct approach to construct exact solution of nonlinear equations.
Once a nonlinear equation is written in bilinear forms by a
dependent variable transformation, then multi-soliton solutions are
usually obtained \cite{Hirota1}-\cite{Sa}.  It was based on Hirota
forms that Nakamura proposed a convenient way to construct a kind of
quasi-periodic solutions of nonlinear equations \cite{Na1, Na2},
where  the periodic  wave solutions of the KdV equation and the
Boussinesq equation were obtained. Such a method indeed exhibits
some advantages. For example, it does not need any Lax pairs and
Riemann surface for the considered equation, allows the explicit
construction of multi-periodic wave solutions, only relies on the
existence of the Hirota's bilinear form, as well as all parameters
appearing in Riemann matrix are arbitrary. Recently, further
development was made to investigate the discrete Toda lattice,
(2+1)-dimensional Kadomtsev-Petviashvili equation  and
Bogoyavlenskii's   breaking soliton equation \cite{Dai}-\cite{Ma2}.
However, where repetitive recursion and computation must be
preformed for each equation \cite{Na1}-\cite{Ma2}.

 The motivation of this paper is to considerably
improves the key steps of the above existing
  methods.  To achieve this aim,  we devise  a theta function bilinear  formula,
which actually provides us a lucid and straightforward  way for
applying in a class of nonlinear  equations. Once a nonlinear
equation is written in bilinear forms, then the double periodic wave
solutions of the nonlinear equation can be obtained directly  by
using the formula. Moreover, we propose a simple and effective
method to analyze asymptotic properties of the periodic solutions.
As illustrative examples, we shall construct double periodic wave
solutions to the shall water wave equation,  (2+1)-dimensional
Bogoyavlenskii-Schiff equation and differential-difference KdV
equation.

  The organization of this paper is as follows. In section 2, we briefly introduce a
 Hirota bilinear operator and a Riemann theta function.
 In particular,  we  provide a key formula for constructing double periodic wave solutions
 for both  differential and difference equations.  As applications of our
method, in sections 3-5,   we construct double periodic wave
solutions to the shall water wave equation, (2+1)-dimensional
Bogoyavlenskii-Schiff equation and differential-difference KdV
equation, respectively.  In addition,  it is rigorously shown that
the double periodic wave solutions tend to the
soliton solutions under small amplitude limits.\\[12pt]
{\bf\large  2.  Hirota bilinear operator and  Riemann theta
function }\\

To fix the notations we recall briefly some notions that will be
used in  this paper. The Hirota bilinear  operators $D_x,  D_t$ and
$D_n$ are defined as follows:
\begin{eqnarray*}
&&D_x^mD_t^k f(x,t)\cdot g(x,
t)=(\partial_x-\partial_{x'})^m(\partial_t-\partial_{t'})^k f(x, t)
g(x', t')|_{x'=x, t'=t}\\
&&e^{\delta D_n}f(n)\cdot
g(n)=e^{\delta(\partial_n-\partial_n')}f(n)g(n')|_{n'=n}=f(n+\delta)g(n-\delta),\\
&&{\rm cosh}(\delta D_n)f(n)\cdot g(n)=\frac{1}{2}(e^{\delta
D_n}+e^{-\delta D_n})f(n)\cdot
g(n),\\
&&{\rm sinh}(\delta D_n) f(n)\cdot g(n)=\frac{1}{2}(e^{\delta
D_n}-e^{-\delta D_n})f(n)\cdot g(n).
\end{eqnarray*}

 {\bf Proposition 1.}   The Hirota
bilinear operators $D_x, D_t$ and $D_n$ have properties
\cite{Hirota1}-\cite{Sa}
\begin{eqnarray*}
&&D_x^mD_t^k   e^{\xi_1}\cdot
  e^{\xi_2}=(\alpha_1-\alpha_2)^m(\omega_1-\omega_2)^k
  e^{\xi_1+\xi_2},\\
  && e^{\delta D_n}  e^{\xi_1}\cdot
  e^{\xi_2}=e^{\delta(\nu_1-\nu_2)}
  e^{\xi_1+\xi_2},\\
  &&{\rm cosh}(\delta D_n) e^{\xi_1}\cdot
  e^{\xi_2}={\rm cosh}[\delta (\nu_1-\nu_2)]e^{\xi_1+\xi_2},\\
  &&{\rm sinh}(\delta D_n) e^{\xi_1}\cdot
  e^{\xi_2}={\rm sinh}[\delta (\nu_1-\nu_2)]e^{\xi_1+\xi_2},
\end{eqnarray*}
where  $\xi_j=\alpha_jx+\omega_jt+\nu_jn+\sigma_j$,  and $\alpha_j,
\ \omega_j, \nu_j,\ \sigma_j$,  $ j=1,2$ are parameters and $n\in
\mathbb{Z}$ is a discrete variable. More generally, we have
$$\begin{aligned}
&F( D_x, D_t, D_n)e^{\xi_1}\cdot
  e^{\xi_2}  =F(\alpha_1-\alpha_2,\omega_1-\omega_2, \exp[\delta(\nu_1-\nu_2)])
  e^{\xi_1+\xi_2},\end{aligned}\eqno(2.1)$$
where $F(D_x, D_t, D_n)$ is a polynomial about operators $ D_x, D_t$
and $ D_n$. This properties  are useful in deriving Hirota's
bilinear form and constructing   periodic wave solutions of
nonlinear equations.

In the following,  we introduce a general  Riemann theta function
and  discuss its periodicity, which plays a central role in the
construction of periodic solutions of nonlinear equations. The
Riemann theta function reads
$$
\vartheta\left[\begin{matrix} \varepsilon\\
s\end{matrix}\right]({\xi},{\tau})=\sum_{{m}\in
\mathbb{{Z}}}\exp\{2\pi i ({\xi}+{\varepsilon})({m}+{s})-\pi {\tau}
({m}+{s})^2\}.\eqno(2.2)$$ Here the integer value ${m}\in
\mathbb{Z}$, complex parameter  ${s}, {\varepsilon}\in \mathcal{C}$,
and complex phase variables ${\xi} \in \mathcal{C}$;
 The ${\tau}>0$  which is
called the period matrix of the Riemann theta function.

  In the definition of the theta function (2.2), for  the
case  ${s}={\varepsilon}={0}$, hereafter we use
$\vartheta({\xi},{{\tau}})=\vartheta\left[\begin{matrix} 0\\
0\end{matrix}\right]({\xi},{\tau})$ for simplicity. Moreover, we
have $\vartheta\left[\begin{matrix} \varepsilon\\
0\end{matrix}\right]({\xi},{\tau})=\vartheta({\xi}+{\varepsilon},
{\tau})$.

{\bf Definition 1.}  A function  $g(t)$ on $\mathbb{C}$  is said to
be quasi-periodic in $t$ with fundamental periods  $T_1, \cdots,
T_k\in \mathbb{C}$,  if
 $T_1, \cdots, T_k$ are linearly dependent over $\mathbb{Z}$ and there exists a function
  $G(y_1, \cdots, y_k)$,  such that
$$ G(y_1,\cdots, y_j+T_j, \cdots, y_k)=G(y_1,\cdots, y_j, \cdots, y_k ),
 \ \ {\rm for\ all}\ y_j\in \mathbb{C}, \ j=1, \cdots, k.$$
$$ G( t,\cdots, t, \cdots, t )=g( t). $$
In particular, $g(t)$ is called double periodic  as $k=2$, and it
becomes periodic with $T$ if and only if $T_j=m_jT, \ j=1, \cdots,
k.$ $\square$

 Let's first see the periodicity of the theta function $\vartheta(\xi, \tau)$.

 {\bf Proposition 2.} \cite{Far} The theta function $\vartheta({\xi}, {\tau})$ has the periodic properties
$$\begin{aligned}
&\vartheta({\xi}+{1}+i{\tau},{\tau})=\exp(-2\pi i\xi+\pi
\tau)\vartheta({\xi},{\tau}).
\end{aligned}\eqno(2.3)$$
 We regard the vectors $1$ and
$i\tau_j$ as periods of the theta function $\vartheta({\xi},{\tau})$
with multipliers $1$ and $\exp({-2\pi i\xi+\pi \tau})$,
respectively.  Here,  $i\tau$  is not a  period  of theta function
$\vartheta({\xi}, {\tau})$,  but it is  the period of the functions
$\partial^2_{\xi}\ln \vartheta({\xi},{\tau})$, $
\partial_{\xi}\ln[\vartheta({\xi}+{e},
{\tau})/\vartheta({\xi}+{h},{\tau})]$ and  $ \vartheta({\xi}+{e},
{\tau})\vartheta({\xi}-{e}, {\tau})/\vartheta({\xi}+{h},{\tau})^2$.

{\bf Proposition 3.} The meromorphic functions $f({\xi})$ on
$\mathbb{C}$ are as follow
$$\begin{aligned}
&(i) \ \ \ \
 \  f({\xi})=\partial_{\xi}^2\ln\vartheta({\xi},
 {\tau}),\ \ {\xi}\in \mathbb{C},\\[3pt]
&(ii) \ \ \ \ f({\xi})=\partial_{\xi}\ln\frac{\vartheta({\xi}+
{e},{\tau})}{ \vartheta({\xi}+{h},{\tau})},\ \ {\xi},\ {e},\ {h}\in
\mathbb{C}.\\[3pt]
&(ii) \ \ \ \ f({\xi})=\frac{\vartheta({\xi}+
{e},{\tau})\vartheta({\xi}- {e},{\tau})}{
\vartheta({\xi},{\tau})^2},\ \ {\xi},\ {e},\ {h}\in \mathbb{C}.
\end{aligned}$$
then in all three cases (i)--(iii), it holds that
$$\begin{aligned}
&f({\xi}+{1}+i{\tau})=f({\xi}), \ \ \ {\xi}\in \mathbb{C},
\end{aligned}\eqno(2.4)$$
that is, $f(\xi)$ is a double periodic function with $1$ and
$i\tau$.

{\it Proof.} By using (2.3),  it is easy to see that
$$\begin{aligned}
&\frac{\partial_\xi\vartheta({\xi}+{1}+i{\tau},{\tau})}
{\vartheta({\xi}+{1}+i{\tau},{\tau})} =-2\pi
i+\frac{\partial_\xi\vartheta({\xi},{\tau})}{\vartheta({\xi},{\tau})},
\end{aligned}$$
or equivalently
$$\begin{aligned}
&\partial_{\xi}\ln\vartheta({\xi}+{1}+i{\tau},{\tau})=-2\pi
i+\partial_{\xi}\ln \vartheta({\xi},{\tau}).
\end{aligned}\eqno(2.5)$$
Differentiating  (2.5) with respective to $\xi$ again immediately
proves the formula (2.4) for the case (i).  The formula (2.4) can be
proved  for the cases (ii) and (iii) in a similar manner.  $\square$

{\bf Theorem 1.}   Suppose that $\vartheta\left[\begin{matrix} \varepsilon'\\
0\end{matrix}\right]({\xi},{\tau})$ and $
\vartheta\left[\begin{matrix} \varepsilon\\
0\end{matrix}\right]({\xi},{\tau})$ are two Riemann theta functions,
in which $\xi=\alpha x+\omega t+\nu n+\sigma$. Then Hirota bilinear
operators $D_x, D_t$ and  $D_n$ exhibit the following perfect
properties when they act on a pair of theta functions
$$\begin{aligned}
&D_x \vartheta\left[\begin{matrix} \varepsilon'\\
0\end{matrix}\right]({\xi},{\tau})\cdot
  \vartheta\left[\begin{matrix} \varepsilon\\
0\end{matrix}\right]({\xi},{\tau})\\
  &=\left[\sum_{{\mu=0,1}}\partial_x\vartheta\left[\begin{matrix} \varepsilon'-\varepsilon\\
-\mu/2\end{matrix}\right](2{\xi}, 2{\tau})|_{{\xi}={0}}\right]
  \vartheta\left[\begin{matrix} \varepsilon'+\varepsilon\\
\mu/2\end{matrix}\right](2{\xi},2{\tau}),
\end{aligned}\eqno(2.6)$$
$$\begin{aligned}
&\exp(\delta D_n) \vartheta\left[\begin{matrix} \varepsilon'\\
0\end{matrix}\right]({\xi},{\tau})\cdot
  \vartheta\left[\begin{matrix} \varepsilon\\
0\end{matrix}\right]({\xi},{\tau})\\
  &=\left[\sum_{{\mu=0,1}}\exp(\delta D_n)\vartheta\left[\begin{matrix} \varepsilon'-\varepsilon\\
-\mu/2\end{matrix}\right](2{\xi}, 2{\tau})|_{{\xi}={0}}\right]
  \vartheta\left[\begin{matrix} \varepsilon'+\varepsilon\\
\mu/2\end{matrix}\right](2{\xi},2{\tau}),
\end{aligned}\eqno(2.7)$$
 where the notation  $\sum_{{\mu=0,1}}$
  represents two different transformations corresponding to  $\mu=0,1$. The bilinear formula for $t$ is the same as (2.6) by replacing
  $\partial_x$ with $\partial_t$.

  In general, for a polynomial operator $F( D_x,
D_t, D_n)$ with respect to  $ D_x, D_t$ and $ D_n$, we have the
following useful formula
$$\begin{aligned}
&F(D_x, D_t, D_n) \vartheta\left[\begin{matrix} \varepsilon'\\
0\end{matrix}\right]({\xi},{\tau})\cdot
  \vartheta\left[\begin{matrix} \varepsilon\\
0\end{matrix}\right]({\xi},{\tau})
  =\left[\sum_{{\mu}}C({\varepsilon'},{\varepsilon}, {\mu})\right]
  \vartheta\left[\begin{matrix} \varepsilon'+\varepsilon\\
\mu/2\end{matrix}\right](2{\xi},2{\tau}),\end{aligned}\eqno(2.8)$$
in which, explicitly
$$\begin{aligned}
  &C({\varepsilon},{\varepsilon'},{\mu})=\sum_{{m}\in \mathbb{Z}^N}
  F({\mathcal{M}})\exp\left[-2\pi{\tau}({m}-{\mu}/2)^2-2\pi i
  ( {m}-{\mu}/2)( {\varepsilon'}-{\varepsilon})\right].
\end{aligned}\eqno(2.9)$$
where we denote vector $ {\mathcal{M}}=(4\pi i(
{m}-{\mu}/2){\alpha},\ 4\pi i(  {m}-{\mu}/2) {\omega}, \exp[4\pi i(
{m}-{\mu}/2)\delta \nu]).$

{\it Proof.} Making use of  Proposition 1, we obtain the relation
\begin{eqnarray*}
&&D_x \vartheta\left[\begin{matrix} \varepsilon'\\
0\end{matrix}\right]({\xi},{\tau})\cdot
  \vartheta\left[\begin{matrix} \varepsilon\\
0\end{matrix}\right]({\xi},{\tau})\\
  &&=\sum_{m',m\in\mathbb{Z}}{D}_x\exp\{2\pi i m'(\xi+\varepsilon')-\pi m'^2{\tau}\}\cdot
  \exp\{2\pi i m(\xi+\varepsilon)-\pi m^2{\tau}\},\\
  &&=\sum_{m',m\in\mathbb{Z}}2\pi i\alpha(m'-m) \exp\left\{2\pi
  i(m'+m)\xi-2\pi i(m'\varepsilon'+m\varepsilon)-\pi{\tau}[m'^2+m^2]\right\}
  \end{eqnarray*}
  By shifting sum index as   $m=l'-m'$, then
  \begin{eqnarray*}
  &&\Delta=\sum_{l',m'\in\mathbb{Z}}2\pi i\alpha(2m'-l') \exp\left\{2\pi
  il'\xi-2\pi i[m'\varepsilon'+(l'-m')\varepsilon] -\pi{\tau}[m'^2+(l'-m')^2]\right\}\\
  &&\stackrel{l'=2l+\mu}{=}\sum_{\mu=0,1}\ \ \sum_{l,m'\in\mathbb{Z}}2\pi i\alpha(2m'-2l-\mu)\exp\{4\pi
  i\xi(l+\mu/2)\\
  &&\ \ \ \ \   -2\pi i[m'\varepsilon'-(m-2l-\mu)\varepsilon] -\pi[m'^2+(m'-2l-\mu)^2]{\tau}\}
  \end{eqnarray*}
Finally letting  $m'=k+l$,  we conclude  that
{\small\begin{eqnarray*}
    &&\Delta=\sum_{\mu=0,1}\left[\sum_{k\in\mathbb{Z}}4\pi i\alpha[k-\mu/2]
    \exp\{-2\pi i(k-\mu/2)(\varepsilon'-\varepsilon)-2\pi{\tau}(k-\mu/2)^2\}\right]\\
&& \ \ \ \ \ \ \ \ \times \left[\sum_{l\in\mathbb{Z}}\exp\{2\pi
i(l+\mu/2)(2\xi+\varepsilon'+\varepsilon)-2\pi{\tau}(l+\mu/2)^2\right]\\
&&=\left[\sum_{{\mu=0,1}}\partial_x\vartheta\left[\begin{matrix} \varepsilon'-\varepsilon\\
-\mu/2\end{matrix}\right](2{\xi}, 2{\tau})|_{{\xi}={0}}\right]
  \vartheta\left[\begin{matrix} \varepsilon'+\varepsilon\\
\mu/2\end{matrix}\right](2{\xi},2{\tau}),
\end{eqnarray*}}
by using  the following relations
\begin{eqnarray*}
    &&k+l=(k-\mu/2)+(l+\mu/2),\ \ k-l-\mu=(k-\mu/2)-(l+\mu/2).
\end{eqnarray*}

In a similar way, we can prove the formula (2.7).  The formula (2.8)
follows from  (2.6) and (2.7). $\Box$

{\bf Remark 1.}  The formulae (2.8) and (2.9) show  that if the
following equations are   satisfied
  $$C({\varepsilon},{\varepsilon'},{\mu})=0,\eqno(2.10)$$
  for   $\mu=0,1$,   then
  $\vartheta\left[\begin{matrix} \varepsilon'\\
0\end{matrix}\right]({\xi},
  {\tau})$ and $
  \vartheta\left[\begin{matrix} \varepsilon\\
0\end{matrix}\right]({\xi},{\tau})$
  are  periodic wave solutions of the bilinear equation
  $$F( D_x, D_t, D_n)\vartheta\left[\begin{matrix} \varepsilon'\\
0\end{matrix}\right]({\xi},
  {\tau})\cdot\vartheta\left[\begin{matrix} \varepsilon\\
0\end{matrix}\right]({\xi},
  {\tau})=0.$$
 The formula (2.10) contains two equations which are called constraint equations.
 This formula  actually provides us an unified approach to construct
  double periodic wave  solutions for both differential and difference equations. Once a
  equation is written bilinear forms, then its  periodic wave  solutions
  can be directly obtained by solving system (2.10).

{\bf Theorem 2.} Let $C({\varepsilon},{\varepsilon'},{\mu})$ and $F(
D_x, D_t, D_n)$  be  given in Theorem 1,  and  make a choice such
that $\varepsilon'-\varepsilon=\pm 1/2$. Then

(i) \ If  $F( D_x, D_t, D_n)$ is an even function in the form
$$F( -D_x, -D_t,-D_n)=F( D_x, D_t, D_n),$$
 then $C({\varepsilon},{\varepsilon'},{\mu})$ vanishes automatically for
 the case  $\mu=1$, namely
  $$C({\varepsilon},{\varepsilon'},{\mu})=0, \ \  {\rm for} \  \ \
 \mu=1.\eqno(2.11)$$

(ii) \ If $F(D_x, D_t, D_n)$ is an odd function in the form
$$F( -D_x,
-D_t, -D_n)=-F( D_x, D_t, D_n),$$ then
$C({\varepsilon},{\varepsilon'},{\mu})$ vanishes automatically for
the case  $\mu=0$, namely
$$C({\varepsilon},{\varepsilon'},{\mu})=0,
\ \ {\rm for} \ \mu=0.\eqno(2.12)$$

{\it Proof.}  We are going to consider the case where  $F(D_x,
D_t,D_n)$ is an even function and prove the formula (2.11). The
formula (2.12) is analogous. Making transformation
${m}=-{\bar{m}}+{\mu}$, and noting $F( D_x, D_t, D_n)$ is even, we
then deduce that
$$\begin{aligned}
  &C({\varepsilon},{\varepsilon'},{\mu})
  =\sum_{{\bar{m}}\in \mathbb{Z}}
  F(-{\mathcal{M}})\exp\left[-2\pi{\tau}({\bar{m}}-{\mu}/2)^2+2\pi i
  ({\bar{m}}-{\mu}/2)( {\varepsilon'}-{\varepsilon})\right]\\
  &\ \ \ \ \ \ \ \ \ \ \ \ \   \ =C({\varepsilon},{\varepsilon'},{\mu})
  \exp\left[4\pi
  i( {\bar{m}}-{\mu}/2)( {\varepsilon'}-{\varepsilon})\right]\\
  &\ \ \ \ \ \ \ \ \ \ \ \ \   \ =C({\varepsilon},{\varepsilon'},{\mu})
  \exp\left(\pm 2\pi i\bar{m}\right)
  \exp\left(\pm\pi i\mu\right) =-C({\varepsilon},{\varepsilon'},{\mu}),
\end{aligned}$$
which proves  the formula (2.11). $\square$

{\bf Corollary 1. }  Let $\varepsilon_j'-\varepsilon_j=\pm 1/2, \
j=1, \cdots, N$.  Assume  $F( D_x, D_t,D_n)$ is a linear combination
of even and  odd functions
$$F( D_x, D_t, D_n)=F_1( D_x, D_t, D_n)+F_2( D_x, D_t, D_n),$$
where $F_1( D_x, D_t, D_n)$ is  even  and $F_2( D_x, D_t, D_n)$ is
odd. In addition, $C({\varepsilon},{\varepsilon'},{\mu})$
corresponding (2.9)  is given by
$$C({\varepsilon},{\varepsilon'},{\mu})
=C_1({\varepsilon},{\varepsilon'},{\mu})
+C_2({\varepsilon},{\varepsilon'},{\mu}),$$ where
  $$C_1({\varepsilon},{\varepsilon'},{\mu})
=\sum_{{m}\in \mathbb{Z}^N}
  F_1({\mathcal{M}})
  \exp\left[-2\pi{\tau}({m}-{\mu}/2)^2-2\pi i
  ( {m}-{\mu}/2)(
  {\varepsilon'}-{\varepsilon})\right],$$
  $$C_2({\varepsilon},{\varepsilon'},{\mu})
=\sum_{{m}\in \mathbb{Z}^N}
  F_2({\mathcal{M}})
  \exp\left[-2\pi{\tau}({m}-{\mu}/2)^2-2\pi i
  ( {m}-{\mu}/2)(
  {\varepsilon'}-{\varepsilon})\right].$$
Then
$$\begin{aligned}
  &C({\varepsilon},{\varepsilon'},{\mu})
  =C_2({\varepsilon},{\varepsilon'},{\mu}) \ \  {\rm for} \  \ \
 \mu=1,
\end{aligned}\eqno(2.13)$$
$$\begin{aligned}
  &C({\varepsilon},{\varepsilon'},{\mu})
  =C_1({\varepsilon},{\varepsilon'},{\mu}), \ \ {\rm for} \ \mu=0.
\end{aligned}\eqno(2.14)$$

{\it Proof.}  In a similar to the proof of Theorem 2,  shifting sum
index as  ${m}=-{\bar{m}}+{\mu}$, and using $F_1( D_x, D_t, D_n)$
even and $F_2(D_x, D_t, D_n)$ odd, we have
$$\begin{aligned}
  &C({\varepsilon},{\varepsilon'},{\mu})
  =C_1({\varepsilon},{\varepsilon'},{\mu})+
  C_2({\varepsilon},{\varepsilon'},{\mu})\\
    &\ \ \ \ \ \ \ \ \ \ \ \ \   \
    =\left[C_1({\varepsilon},{\varepsilon'},{\mu})-
  C_2({\varepsilon},{\varepsilon'},{\mu})\right]
  \exp\left(\pm\pi i\mu\right).
\end{aligned}\eqno(2.15)$$

Then  for  $\mu=1$,  the equation (2.15) gives
$$C_1({\varepsilon},{\varepsilon'},{\mu})=0,$$
which implies the formula (2.13). The formula (2.14) is is
analogous. $\square$

The theorem 2 and corollary 1 are very useful to deal with coupled
Hirota's bilinear equations, which will  be  seen  in the following
section 4.
\\[12pt]
{\bf\large 3.   The shall water wave  equation}\\

The  shall water wave  equation takes the form \cite{Clarkson}
$$\begin{aligned}
&u_t-u_{xxt}-3uu_t+3u_x\int_x^{\infty}u_tdx+u_x=0,
\end{aligned}\eqno(3.1)$$
 which is like to the KdV equation in the family of shall water wave equations.
 Hirota and Satsuma obtained soliton solutions of the equation by means of
 bilinear method  \cite{Satsuma}. Here we construct its a double periodic wave solution
 and show that the one-soliton solution can be obtained as limiting case of
 the double periodic solution.

 To apply the Hirota bilinear method  for constructing
double periodic wave solutions of the equation (3.1),  we   consider
a
 variable transformation
$$u=2\partial_x^2 \ln f(x, t).\eqno(3.2)$$
 Substituting (3.2) into (3.1) and integrating with respect to $x$,
we then get the following Hirota's bilinear form
$$\begin{aligned}
&F(D_x,D_t)f\cdot f=(D_xD_t+D_x^2-D_tD_x^3+c) f\cdot f=0,
\end{aligned}\eqno(3.3)$$ where
 $c$ is an
integration constant. In the special case of $c=0$, starting from
the bilinear equation (3.3), it is easy to find  its one-soliton
solution
$$u_1=2\partial_x^2\ln(1+e^{\eta}),\eqno(3.4)$$
with  phase variable $\eta= p x+\frac{p}{p^2-1} t+\gamma$ for every
$p$  and $\gamma$.

Next, we turn to see the periodicity of the solution (3.2), the
function  $f$ is chosen to be a Riemann theta function, namely,
 $$f(x,t)=\vartheta({\xi}, {\tau}),\eqno(3.5)$$
where  phase variable $\xi=\alpha x+\omega t+\sigma.$   With
Proposition 3, we refer to
$$u=2\partial_{x}^2\ln\vartheta({\xi},{\tau})=2\alpha^2\partial_{\xi}^2\ln\vartheta({\xi},{\tau}),\eqno(3.6)$$
which shows that  the solution  $u$ is  a double periodic function
with two fundamental periods $1$ and $i{\tau}$.

We introduce the notations by
$$\begin{aligned}
&\lambda=e^{-\pi\tau/2 },\quad
\vartheta_1(\xi,\lambda)=\vartheta(2\mathbf{\xi},2\tau)=\sum_{m\in\mathbb{Z}}
\lambda^{4m^2}\exp(4i\pi m \xi),\\
&\vartheta_2(\xi,\lambda)=\vartheta\left[\begin{matrix} 0\\
-1/2\end{matrix}\right](2\mathbf{\xi},2\tau)=\sum_{m\in\mathbb{Z}}
\lambda^{(2m-1)^2}\exp[2i\pi(2m-1) \xi],\end{aligned}\eqno(3.7)$$
where the phase variable $\xi=\alpha x+\omega t+\sigma$.

Substituting (3.5) into (3.3),  using formula (2.10) and (3.7) leads
to a linear system ( corresponding to $\mu=0$ and $\mu=1$,
respectively)
 $$\begin{aligned}
&[\vartheta_1''(0,\lambda)\alpha+\vartheta_1^{(4)}(0,\lambda)\alpha^4]\omega+\vartheta_1(0,\lambda)c+
\vartheta_1''(0,\lambda)\alpha^2=0,\\
&[\vartheta_2''(0,\lambda)\alpha+\vartheta_2^{(4)}(0,\lambda)\alpha^4]\omega+\vartheta_2(0,\lambda)c+
\vartheta_2''(0,\lambda)\alpha^2=0,
\end{aligned}\eqno(3.8)$$
where we have  denoted the derivative of $\vartheta_j(\xi,\lambda)$
at $\xi=0$  by notations
$$\vartheta_j^{(k)}(0,\lambda)=\frac{d^k\vartheta_j(\xi,\lambda)}{d\xi^k}|_{\xi=0}, \ \ j=1,2; k=1, 2, 3, 4.$$
  This system
admits an explicit solution  $(\omega, c)$. In this way, we obtain
an explicit periodic wave solution (3.6) with parameters $\omega$,
$c$ by (3.8), while other parameters $\alpha, \sigma, \tau, \sigma$
are free.

 In summary,  double periodic
wave (3.6) possesses the following features:
 (i) It is is one-dimensional, i.e. there is a single phase
variable $\xi$. Moreover, it  has two fundamental periods $1$ and
$i\tau$ in phase variable $\xi$, but it need not to be periodic in
$x$ and $t$ . (ii) It can be viewed as a parallel superposition of
overlapping one-soliton waves, placed one period apart.

In the following, we further consider asymptotic properties of the
periodic wave solution. Interestingly,  the relation between the
periodic wave solution (3.6) and the one-soliton solution (3.4) can
be   established  as follows.

{\bf Theorem 3.}  Suppose that the vector  $(\omega, c)$ is a
solution of the system (3.8), and for the periodic wave solution
(3.6), we let
$$ \alpha=\frac{p}{2\pi i}, \ \  \sigma=\frac{\gamma+\pi \tau}{2\pi i},\eqno(3.9)$$ where
the $p$ and $\gamma$ are given  in (3.4). Then we have the following
asymptotic properties
$$
\begin{aligned}
&c\longrightarrow 0, \ \  2\pi
i\xi-\pi\tau\longrightarrow\eta=px+\frac{p}{p^2-1}t
+\gamma,\\
&\vartheta(\xi,\tau)\longrightarrow 1+e^{\eta}, \ \ {\rm as } \ \
\lambda\rightarrow 0.
 \end{aligned}
 $$
 In
other words, the double periodic solution (3.10)  tends to the
soliton solution (3.4) under a small amplitude limit, that is,
$$u\longrightarrow u_1, \ \ {\rm as } \ \
\lambda\rightarrow 0.\eqno(3.10)$$

{\it Proof.}  Here we will directly use the system (3.8) to analyze
asymptotic properties of the periodic solution (3.6).  Since the
coefficients of system (3.8) are power series about $\lambda$, its
solution $(\omega, c)$ also should be a series about $\lambda$.  We
explicitly expand the coefficients of system (3.8) as follows
$$\begin{aligned}
&\vartheta_1(0,\lambda)=1+2\lambda^4+\cdots,\quad
\vartheta_1''(0,\lambda)=-32\pi^2\lambda^{4}+\cdots,\\
&\vartheta_1^{(4)}(0,\lambda)=512\pi^4\lambda^4+\cdots,
\ \ \vartheta_2(0,\lambda)=2\lambda+2\lambda^9+\cdots\\
&\vartheta_2''(0,\lambda)=-8\pi^2\lambda+\cdots,\
\vartheta_2^{(4)}(0,\lambda)
=32\pi^4\lambda+\cdots.\end{aligned}\eqno(3.11)$$ Let  the solution
of the system (3.8) be in the form
$$\begin{aligned}
&\omega=\omega_0+\omega_1\lambda+\omega_2\lambda^2+\cdots=\omega_0+o(\lambda),\\
&c=c_0+c_1\lambda+c_2\lambda^2+\cdots=c_0+o(\lambda).
\end{aligned}\eqno(3.12)$$

Substituting the expansions (3.11) and (3.12) into the system (3.8)
(the second equation is divided by $\lambda$ ) and letting
$\lambda\longrightarrow 0$, we immediately obtain
 the following relations
$$
 \begin{aligned}
 &c_0=0, \ \ (-8\pi^2\alpha+32\pi^4\alpha^3)\omega_0-8\pi^2\alpha^2=0,
 \end{aligned}$$
which implies
 $$c_0=0, \ \ w_0=\frac{\alpha}{4\pi^2\alpha^2-1}.\eqno(3.13)$$
Combining (3.12) and (3.13) then  yields
$$c\longrightarrow 0, \ \  2\pi i\omega\longrightarrow \frac{2\pi i\alpha}{(2\pi i\alpha)^2-1}=
\frac{p}{p^2-1}, \ \ {\rm as } \ \ \lambda\rightarrow 0.$$ Hence we
conclude
$$\begin{aligned}
&\hat{\xi}=2\pi i\xi-\pi \tau=p x+2\pi i\omega t+\gamma\\
&\quad \longrightarrow px+\frac{p}{p^2-1}t+\gamma=\eta,\ \ {\rm as}\
\ \lambda\rightarrow 0.
\end{aligned}\eqno(3.14)$$

 It remains to consider  asymptotic properties of  the periodic wave solution (3.6) under the limit
$\lambda\rightarrow 0$. By expanding the Riemann theta function
$\vartheta(\xi, \tau)$ and using (3.14), it follows that
$$\begin{aligned}
&\vartheta(\xi,\tau)=1+\lambda^2(e^{2\pi i\xi}+e^{-2\pi
i\xi})+\lambda^8(e^{4\pi i\xi}+e^{-4\pi i\xi}) +\cdots
\\
&\ \ \ \
 \ =1+e^{\hat{\xi}}+\lambda^4(e^{-\hat{\xi}}+e^{2\hat{\xi}})+\lambda^{12}(e^{-2\hat{\xi}}+e^{3\hat{\xi}})
+\cdots\\
&\ \ \ \
 \ \quad \longrightarrow 1+e^{\hat{\xi}}\longrightarrow 1+e^{\eta},\ \
{\rm as}\ \ \lambda\rightarrow 0,
\end{aligned}$$
 which together with (3.6) leads to (3.10). Therefore we conclude that the double  periodic solution
(3.6) just goes to the one-soliton solution (3.4)  as the amplitude
$\lambda\rightarrow 0$. $\square$\\[12pt]
{\bf\large 4. The modified Bogoyavlenskii-Schiff equation}\\

We consider (2+1)-dimensional modified Bogoyavlenskii-Schiff
equation \cite{ Yu}
$$\begin{aligned}
&u_t-4u^2u_z-2u_x\partial_x^{-1}(u^2)_z+u_{xxz}=0,
\end{aligned}\eqno(4.1)$$
which was deduced from the Miura transformation
\cite{Bogoyavenskii}.  Equation (4.1) is reduced to the modified KdV
equation in the case of $x=z$.

  We shall construct a double periodic wave solution to the equation
(4.1) by using Theorem 1 and 2.  The  equation  (4.1) can be
described by a coupled system
$$\begin{aligned}
&u=\psi_x, \\
& \rho_{xx}+\psi_x^2+c=0,\\
&\psi_t+2\psi_x\rho_{xz}+\psi_z(\rho_{xx}+\psi_x^2+c)+\psi_{xxz}=0.
\end{aligned}\eqno(4.2)$$
 We perform the dependent variable transformations
$$\begin{aligned}
&u=\psi_x=\partial_x\ln\left(\frac{f}{g}\right),\ \ \rho=\ln(fg),
\end{aligned}\eqno(4.3)$$
then equation (4.2) is reduced to the following bilinear form
$$\begin{aligned}
&F(  D_x)f\cdot
g=(D_x^2+c) f\cdot g=0,\\
&G( D_t, D_x, D_z)f\cdot g=(D_t+D_x^2D_z+cD_z) f\cdot g=0,
\end{aligned}\eqno(4.4)$$ where
 $c$ is a constant. The equation (4.4)  is
 a  type of coupled bilinear equations, which is more difficult to be dealt with than
 the single bilinear equation (3.3) due to appearance of two functions and two equations. We will take full  advantages of
 Theorem 2 to reduce the number of constraint  equations.

Now we take into account  the periodicity of the solution (4.3), in
which we take
 $f$  and $g$   as
 $$f=\vartheta({\xi}+e, {\tau}),
 \ \ g=\vartheta({\xi}+h, {\tau}), \ \ {e},{h}\in \mathbb{C},
\eqno(4.5)$$
 where  phase variable $\xi=\alpha x+\beta z+\omega
t+\sigma.$ By means of Proposition 3, we find that the solution
$$\begin{aligned}
&u=\alpha \partial_{\xi}\ln\frac{\vartheta({\xi}+e, {\tau})}
{\vartheta({\xi}+h, {\tau})}
\end{aligned}$$
 is a  double periodic function
with two fundamental periods $1$ and $i\tau$.

 In the special case of $c=0$,   the
equation (4.2) admits one-soliton solution
$$u_1=\partial_x\ln\frac{1+e^{\eta}}{1-e^{\eta}},\eqno(4.6)$$
where  $\eta= p x+qy-p^2q t+\gamma$ for every $p, q$ and $\gamma$.

 We  take $e=0, \ h=1/2$ in (4.5), and therefore
$$\begin{aligned}
&f=\vartheta(\xi,\tau)=\sum_{m\in\mathbb{Z}}\exp({2\pi
in\xi-\pi m^2\tau}),\\
&g=\vartheta\left[\begin{matrix} 1/2\\
0\end{matrix}\right](\xi,\tau)=\sum_{m\in\mathbb{Z}}\exp({2\pi
im(\xi+1/2)-\pi m^2\tau})\\
&\ \ \ =\sum_{m\in\mathbb{Z}}(-1)^m\exp({2\pi im\xi-\pi
m^2\tau}).\end{aligned}\eqno(4.7)$$

Due to the fact that  $F( D_x)$ is an even function, its constraint
equations in the formula (2.10)  vanish automatically for $\mu=1$.
Similarly the constraint equations associated with $G(D_t, D_x,
D_z)$ also vanish automatically for $\mu=0$. Therefore, the Riemann
theta function (4.6) is a solution of the bilinear equation (4.4),
provided the following equations
$$\begin{aligned}
&\vartheta_1''(0,\lambda)\alpha^2+\vartheta_1(0,\lambda)c=0,\\
&\vartheta_2'(0,\lambda)\omega+\vartheta_2'(0,\lambda)\beta c  +
\vartheta_2'''(0,\lambda)\alpha^2\beta=0,
\end{aligned}\eqno(4.8)$$
where we introduce the notations by
$$\begin{aligned}
&\lambda=e^{-\pi\tau/2 },\quad
\vartheta_1(\xi,\lambda)=\vartheta(2\mathbf{\xi},2\tau)=\sum_{m\in\mathbb{Z}}
\lambda^{4m^2}\exp(4i\pi m \xi),\\
&\vartheta_2(\xi,\lambda)=\vartheta\left[\begin{matrix} 1/2\\
-1/2\end{matrix}\right](2\mathbf{\xi},2\tau)=\sum_{m\in\mathbb{Z}}
(-1)^m\lambda^{(2m-1)^2}\exp[2i\pi(2m-1) \xi].\end{aligned}$$ It is
obvious that equation (4.8) admits  an explicit solution $\omega$
and $c$. In this way,  a periodic wave solution reads
$$u=\partial_x\ln\frac{\vartheta(\xi,\tau)}{\vartheta(\xi+1/2,\tau)},\eqno(4.9)$$
 where parameters $\omega$ and $c$ are given
by (4.11),  while  other parameters $\alpha, \beta, \tau, \sigma$
are free. In summary, double periodic wave (4.9)  has the following
features: (i) It is one-dimensional and has two fundamental periods
$1$ and $i\tau$  in phase variable $\xi$.  (ii) It can be viewed as
a parallel superposition of overlapping one-soliton waves, placed
one period apart.

In the following, we further consider asymptotic properties of the
 double periodic wave solution. The relation between the periodic wave
solution (4.9) and the one-soliton solution (4.6) can be established
as follows.

{\bf Theorem 4.}  Suppose that the vector  $(\omega, c)^T$ is a
solution of the system (4.8). In the periodic wave solution (4.9),
we choose parameters as
$$ \alpha=\frac{p}{2\pi i}, \ \ \beta=\frac{q}{2\pi
i},\ \ \sigma=\frac{\gamma+\pi \tau}{2\pi i},\eqno(4.10)$$ where the
$p, q$ and $\gamma$ are the same as those  in (4.6). Then we have
the following asymptotic properties
$$c\longrightarrow 0, \ \  \xi\longrightarrow\frac{\eta+\pi\tau}{2\pi i}, \ \
f\longrightarrow 1+e^{\eta}, \ \ g\longrightarrow 1-e^{\eta}, \ \
{\rm as } \ \ \lambda\rightarrow 0.$$   In other words, the double
periodic solution (4.9) tends to the one-soliton solution (4.6)
 under a small amplitude limit , that is,
$$u\longrightarrow u_1, \ \ {\rm as } \ \
\lambda\rightarrow 0.\eqno(4.11)$$

{\it Proof.}  Here we will directly use the system (4.8) to analyze
asymptotic properties of periodic solution (4.9).  We explicitly
expand the coefficients of system (4.8) as follows
$$\begin{aligned}
&\vartheta_1(0,\lambda)=1+2\lambda^4+\cdots,\quad
\vartheta_1''(0,\lambda)=-32\pi^2\lambda^{4}+\cdots,\\
&\vartheta_2'(0,\lambda)=-4\pi i\lambda+12\pi i\lambda^9+\cdots, \ \
\vartheta_2'''(0,\lambda)=16\pi^3i\lambda-48\pi^3i\lambda^9+\cdots,\end{aligned}\eqno(4.12)$$
Suppose that  the solution of the system (4.8) is of the form
$$\begin{aligned}
&\omega=\omega_0+\omega_1\lambda+\omega_2\lambda^2+\cdots=\omega_0+o(\lambda),\\
&c=c_0+c_1\lambda+c_2\lambda^2+\cdots=c_0+o(\lambda).
\end{aligned}\eqno(4.13)$$

Substituting the expansions (4.12) and (4.13) into the system (4.8)
and letting $\lambda\longrightarrow 0$, we immediately obtain
 the following relations
$$
 \begin{aligned}
 &c_0=0,\ \ -4\pi i\omega_0+16\pi^3i\alpha^2\beta=0,\ \
 \end{aligned}$$
which has a solution
 $$c_0=0, \ \ w_0=4\pi^2\alpha^2\beta.\eqno(4.14)$$
Combining  (4.13) and (4.14) leads to
$$c\longrightarrow 0, \ \  2\pi i\omega\longrightarrow 8\pi^3i\alpha^2\beta=-p^2q, \ \ {\rm as } \ \
\lambda\rightarrow 0,$$ or equivalently
$$\begin{aligned}
&\hat{\xi}=2\pi i\xi-\pi \tau=p x+qy+2\pi i\omega t+\gamma\\
&\quad \longrightarrow px+qy-p^2qt+\gamma=\eta,\ \ {\rm as}\ \
\lambda\rightarrow 0.
\end{aligned}\eqno(4.15)$$

 It remains to  identify   that the periodic wave  (4.9) possesses  the same
form with the one-soliton solution (4.6) under the limit
$\lambda\rightarrow 0$. For this purpose, we start to  expand the
 functions $f$  and $g$ in the form
$$ f=1+\lambda^2(e^{2\pi i\xi}+e^{-2\pi i\xi})+\lambda^8(e^{4\pi i\xi}+e^{-4\pi i\xi})
+\cdots .$$
$$ g=1-\lambda^2(e^{2\pi i\xi}+e^{-2\pi i\xi})+\lambda^8(e^{4\pi i\xi}+e^{-4\pi i\xi})
+\cdots .$$
 By using   (4.13)-(4.15),  it follows that
$$\begin{aligned}
&f=1+e^{\hat{\xi}}+\lambda^4(e^{-\hat{\xi}}+e^{2\hat{\xi}})+\lambda^{12}(e^{-2\hat{\xi}}+e^{3\hat{\xi}})
+\cdots\\
&\quad \longrightarrow 1+e^{\hat{\xi}}\longrightarrow 1+e^{\eta},\ \
{\rm as}\ \ \lambda\rightarrow 0;\\
&g=1-e^{\hat{\xi}}+\lambda^4(e^{2\hat{\xi}}-e^{-\hat{\xi}})+\lambda^{12}(e^{-2\hat{\xi}}-e^{3\hat{\xi}})
+\cdots\\
&\quad \longrightarrow 1-e^{\hat{\xi}}\longrightarrow 1-e^{\eta},\ \
{\rm as}\ \ \lambda\rightarrow 0.
\end{aligned}\eqno(4.16)$$
The expression (4.11) follows from  (4.16), and thus we conclude
that the double  periodic solution (4.9) just goes to the
one-soliton solution (4.6) as the amplitude
$\lambda\rightarrow 0$. $\square$\\[12pt]
{\bf\large 5.  The differential-difference KdV equation}\\

We consider differential-difference KdV equation
$$\begin{aligned}
&\frac{d}{dt}\left(\frac{u(n)}{1+u(n)}\right)=u(n-1/2)-u(n+1/2).
\end{aligned}\eqno(5.1)$$
Hirota and Hu have found its soliton solutions and rational
solutions  \cite{Hirota2, Hu2},  among them   one-soliton solution
reads
$$u_{1}(n)=\frac{(1+e^{\eta+p/2})(1+e^{\eta-p/2})} {(1+e^{\eta})^2}-1,\eqno(5.2)$$
where  $\eta=pn-\sinh(p/2)t+\gamma$ for every $p$ and $\gamma$.

  We shall construct a periodic wave solutions to the equation (5.1) by using Theorem 1. By
means of a variable transformation
$$\begin{aligned}
&u(n)=\frac{f({n+1/2}) f({n-1/2})}{f(n)^2}-1,
\end{aligned}\eqno(5.3)$$
the equation (5.1) is reduced to the bilinear equation
$$\begin{aligned}
&\left[\sinh(\frac{1}{4}D_n)D_t+2\sinh(\frac{1}{4}D_n)\sinh(\frac{1}{2}D_n)+c\right]f(n)\cdot
f(n)=0,
\end{aligned}\eqno(5.4)$$
where $c$ is a constant.

 Now we take into account the periodicity of the solution
(5.3), in which we take $f(n)=\vartheta({\xi}, {\tau}), $ where
phase variable $\xi=\nu n+\omega t+\sigma.$ Then solution (5.3) is
written as
$$\begin{aligned}
&u(\xi)\equiv u(n)=\frac{\vartheta({\xi}+\frac{1}{2}\nu,
{\tau})\vartheta({\xi}-\frac{1}{2}\nu, {\tau})} {\vartheta({\xi},
{\tau})^2}-1.
\end{aligned}\eqno(5.5)$$
By means of Proposition 2, it is easy to deduce that $u_n$ is a
double periodic function with two fundamental periods $1$ and
$i\tau$.

 Substituting (5.5) into (5.4) and using formula (2.10) leads to a
 linear system
 $$\begin{aligned}
&\sinh(\frac{1}{4}D_n)\vartheta_1'(0,\lambda)\omega+\vartheta_1(0,\lambda)c+
\sinh(\frac{1}{4}D_n)\sinh(\frac{1}{2}D_n)\vartheta_1(0,\lambda)=0,\\
&\sinh(\frac{1}{4}D_n)\vartheta_2'(0,\lambda)\omega+\vartheta_2(0,\lambda)c+
\sinh(\frac{1}{4}D_n)\sinh(\frac{1}{2}D_n)\vartheta_2(0,\lambda)=0,
\end{aligned}\eqno(5.6)$$
where $\vartheta_1(\xi,\lambda)$ and $\vartheta_2(\xi,\lambda)$ are
the same as those in (3.7) with $\xi=\nu n+\omega t+\sigma.$ By
using the solution $\omega$ and $c$ of system (5.6),  a periodic
wave solution is obtained by (5.5).

In the following, we further consider asymptotic properties of the
double periodic wave solution. The relation between the periodic
wave solution (5.5) and the one-soliton solution (5.2) can be
established  as follows.

{\bf Theorem 5.}  Suppose that the vector  $(\omega, c)^T$ is a
solution of the system (5.6). In the periodic wave solution (5.5),
we choose parameters as
$$ \nu=\frac{p}{2\pi i}, \ \ \sigma=\frac{\gamma+\pi \tau}{2\pi i},\eqno(5.7)$$ where the
$p$ and $\gamma$ are the same as those  in (5.2). Then we have the
following asymptotic properties
$$c\longrightarrow 0, \ \  \xi\longrightarrow\frac{\eta+\pi\tau}{2\pi i}, \ \
\vartheta(\xi, \tau)\longrightarrow 1+e^{\eta},  \ \ {\rm as } \ \
\lambda\rightarrow 0.$$   In other words, the periodic solution
(5.5) tends to the one-soliton solution (5.2)
 under a small amplitude limit , that is,
$$u(n)\longrightarrow u_{1}(n), \ \ {\rm as } \ \
\lambda\rightarrow 0.\eqno(5.8)$$

{\it Proof.}  Here we will directly use the system (5.6) to analyze
asymptotic properties of periodic solution (5.5).  We explicitly
expand the coefficients of system (5.6) as follows
$$\begin{aligned}
&\vartheta_1(0,\lambda)=1+2\lambda^4+\cdots,\quad
\sinh(\frac{1}{4}D_n)\vartheta_1'(0,\lambda)=8\pi i\sinh(i\pi\nu)\lambda^{4}+\cdots,\\
&\sinh(\frac{1}{4}D_n)\sinh(\frac{1}{2}D_n)\vartheta_1(0,\lambda)=2\sinh(i\pi\nu)\sinh(2i\pi\nu)\lambda^{4}+\cdots,\\
&\vartheta_2(0,\lambda)=2\lambda+2\lambda^9+\cdots, \ \
\sinh(\frac{1}{4}D_n)\vartheta_2'(0,\lambda)=4\pi i\sinh(i\pi\nu/2)\lambda+\cdots,\\
&\sinh(\frac{1}{4}D_n)\sinh(\frac{1}{2}D_n)\vartheta_2(0,\lambda)=2\sinh(i\pi\nu)\sinh(i\pi\nu/2)\lambda+\cdots.
\end{aligned}\eqno(5.9)$$
Suppose that  the solution of the system (5.6) is of the form
$$\begin{aligned}
&\omega=\omega_0+\omega_1\lambda+\omega_2\lambda^2+\cdots=\omega_0+o(\lambda),\\
&c=c_0+c_1\lambda+c_2\lambda^2+\cdots=c_0+o(\lambda).
\end{aligned}\eqno(5.10)$$

Substituting the expansions (5.9) and (5.10) into the system (5.6)
and letting $\lambda\longrightarrow 0$, we immediately obtain
 the following relations
$$
 \begin{aligned}
 &c_0=0,\ \ 4\pi i\sinh(i\pi\nu/2)\omega_0+2\sinh(i\pi\nu/2)\sinh(i\pi\nu)=0,\ \
 \end{aligned}$$
which implies
 $$c_0=0, \ \ w_0=-\frac{1}{2\pi i}\sinh(i\pi\nu).\eqno(5.11)$$
Combining  (5.9) and (5.10) leads to
$$c\longrightarrow 0, \ \  2\pi i\omega\longrightarrow -\sinh(i\pi \nu)=-\sinh(p/2), \ \ {\rm as } \ \
\lambda\rightarrow 0,$$ or equivalently
$$\begin{aligned}
&\hat{\xi}=2\pi i\xi-\pi \tau=p n+2\pi i\omega t+\gamma\\
&\quad \longrightarrow pn-\sinh(p/2)t+\gamma=\eta,\ \ {\rm as}\ \
\lambda\rightarrow 0.
\end{aligned}\eqno(5.12)$$

  It remains to consider  asymptotic properties of  the periodic wave solution (5.5) under the limit
$\lambda\rightarrow 0$. By expanding the Riemann theta function
$\vartheta(\xi, \tau)$, it follows that
$$\begin{aligned}
&\vartheta(\xi,\tau)=1+e^{\hat{\xi}}+\lambda^4(e^{-\hat{\xi}}+e^{2\hat{\xi}})+\lambda^{12}(e^{-2\hat{\xi}}+e^{3\hat{\xi}})
+\cdots\\
&\ \ \ \
 \ \quad \longrightarrow 1+e^{\hat{\xi}}\longrightarrow 1+e^{\eta},\ \
{\rm as}\ \ \lambda\rightarrow 0,
\end{aligned}$$
 which together with (5.5) lead to (5.8). Therefore we conclude that the periodic solution
(5.5) just goes to the one-soliton solution (5.2)  as the amplitude
$\lambda\rightarrow 0$. $\square$\\[12pt]
{\bf\large  Acknowledgment}

  The work  described in this paper was supported by grants from Research Grants Council contracts HKU,
   the National Science Foundation of China (No.10971031),
Shanghai Shuguang Tracking Project (No.08GG01) and Innovation
Program of Shanghai Municipal Education Commission (No.10ZZ131).

\end{document}